# Dynamic transitions of blind spots in the Hermann grid illusion

Yutaka Nishiyama

Osaka University of Economics,

2, Osumi Higashiyodogawa Osaka, 533-8533, Japan

http://yutaka-nishiyama.sakura.ne.jp/index.html

nishiyama@osaka-ue.ac.jp

## Abstract

Hermann discovered the grid illusion in 1870, but its cause has remained a mystery for more than 150 years. In 1960, Baumgartner proposed a hypothesis for the illusion based on neural receptive fields, but Geier presented a counterexample in 2008. In 1995, Schrauf devised the scintillating grid illusion, an improvement on the Hermann grid illusion. I propose that a hypothesis involving blind spots (optic discs) can significantly contribute to unraveling the mystery of the grid illusion.

## 1. The Hermann grid illusion

In 1870, German physiologist Ludimar Hermann (1838–1914) discovered the grid illusion in 1870 while reading a German translation of a lecture transcript by physicist John Tyndall (1820–1893)[1]. On page 169 of that manuscript was a figure related to sound, consisting of forty black squares neatly arranged into eight rows and five columns, each containing a Chladni figure representing sound vibrations[2]. Hermann noticed faint shadows appearing at the intersections of the white grid formed between the black squares. However, no such shadows were actually printed. Interestingly, Hermann became fascinated not with the sounds represented in the figure, but with the optical illusion it produced.

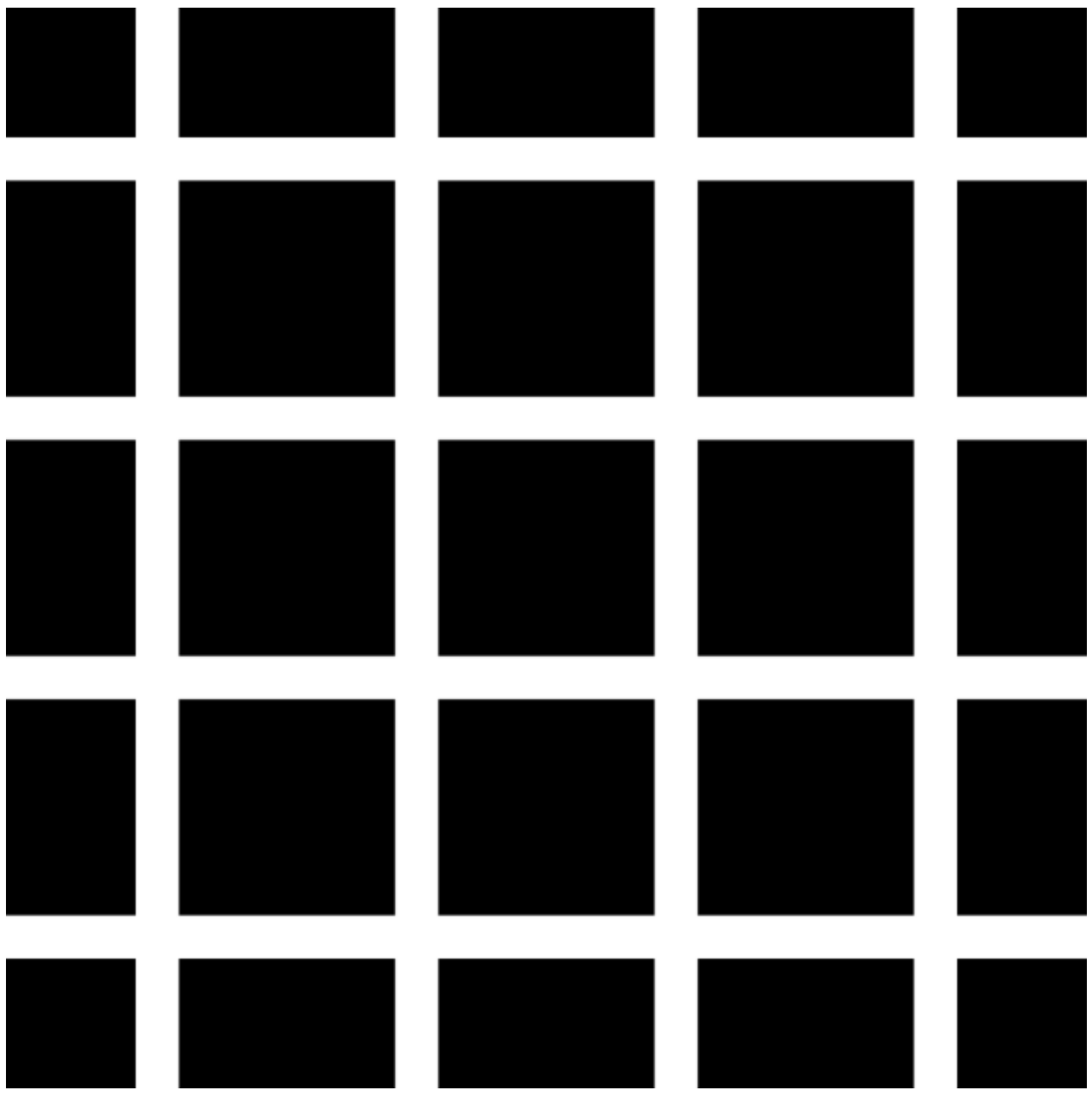

Figure 1: The Hermann grid illusion (1870)[1].

Focusing your gaze at the center of Figure 1 while directing attention to the surrounding intersections, you can see faint, dark shadows. If you try to look directly at these shadows, however, they disappear; they don't actually exist in the image.

## 2. Baumgartner's lateral inhibition and a counterexample

The common explanation for the Hermann grid illusion is Baumgartner's theory[3], which involves central and surrounding neurons. Figure 2 compares the receptive fields of neurons at intersections and along the paths. Around the intersections, there are white areas in both vertical and horizontal directions, while along the paths, white areas are only in either a vertical or horizontal direction. In other words, the area around intersections is brighter than the area around paths. In the figure, plus signs represent bright areas, and minus signs represent dark areas.

One might predict different responses from the central and surrounding neurons. Namely, the response to intersection centers would likely be weaker than the response to path centers because in the former case, the neural output would be suppressed since the surroundings are less dark. This is called lateral inhibition of neurons. As a result, the intersection centers appear darker than other parts of the path.

However, the dark shadows do not appear at directly observed intersections. This is thought to be because at the point of focus, i.e., the center of the visual field, the neurons' receptive fields are very small. Consequently, there is no difference in neural response between intersection and path centers.

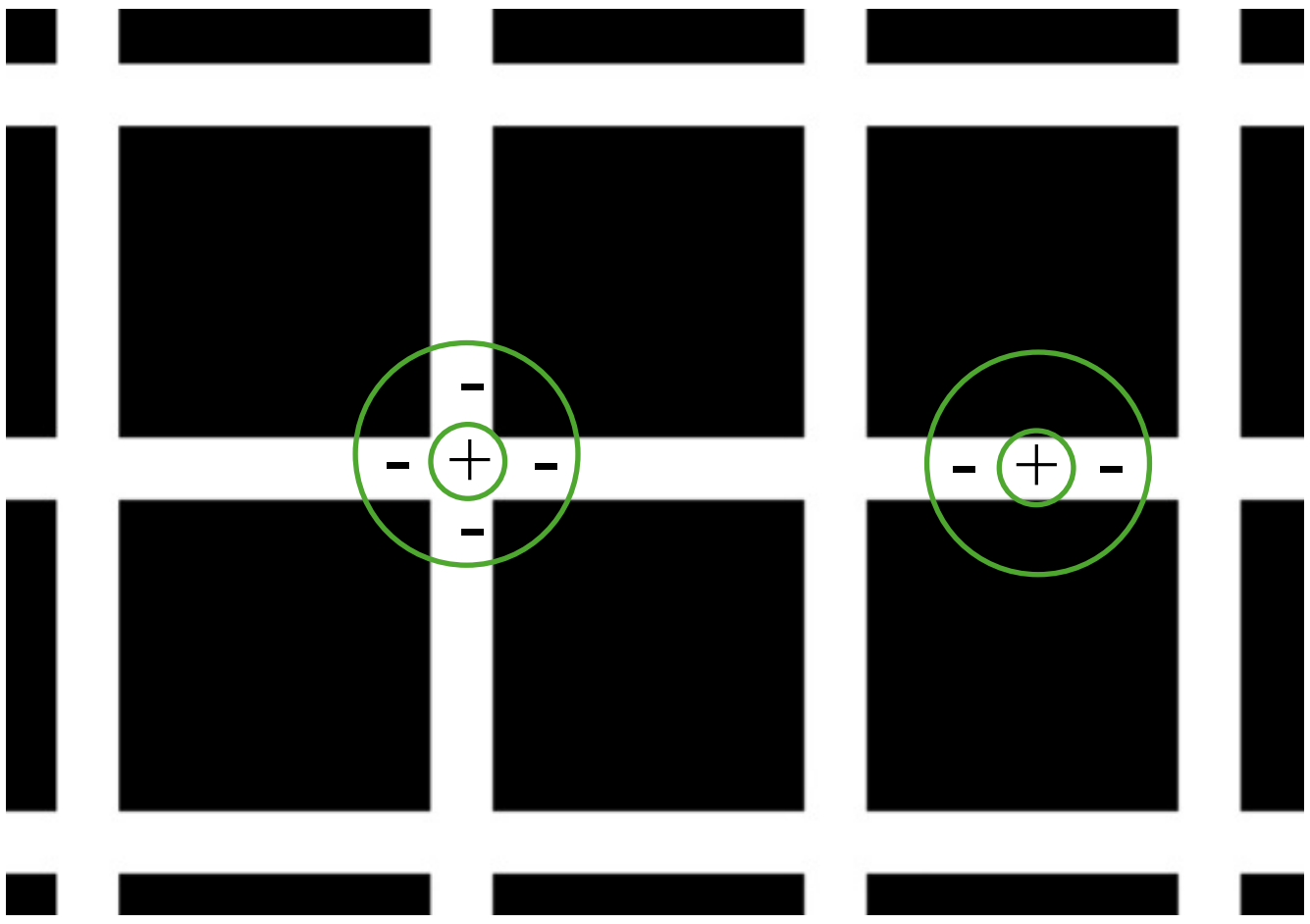

Figure 2: Baumgartner's receptive fields (1960)[3].

However, researchers have discovered several phenomena that lateral inhibition of neurons cannot explain. For example, shadows do not appear at the intersections in Geier's pattern (Figure 3), which distorts the paths in the Hermann grid illusion[4)5)]. Baumgartner's hypothesis of central and surrounding neurons cannot explain this, which suggests that a different mechanism is taking effect.

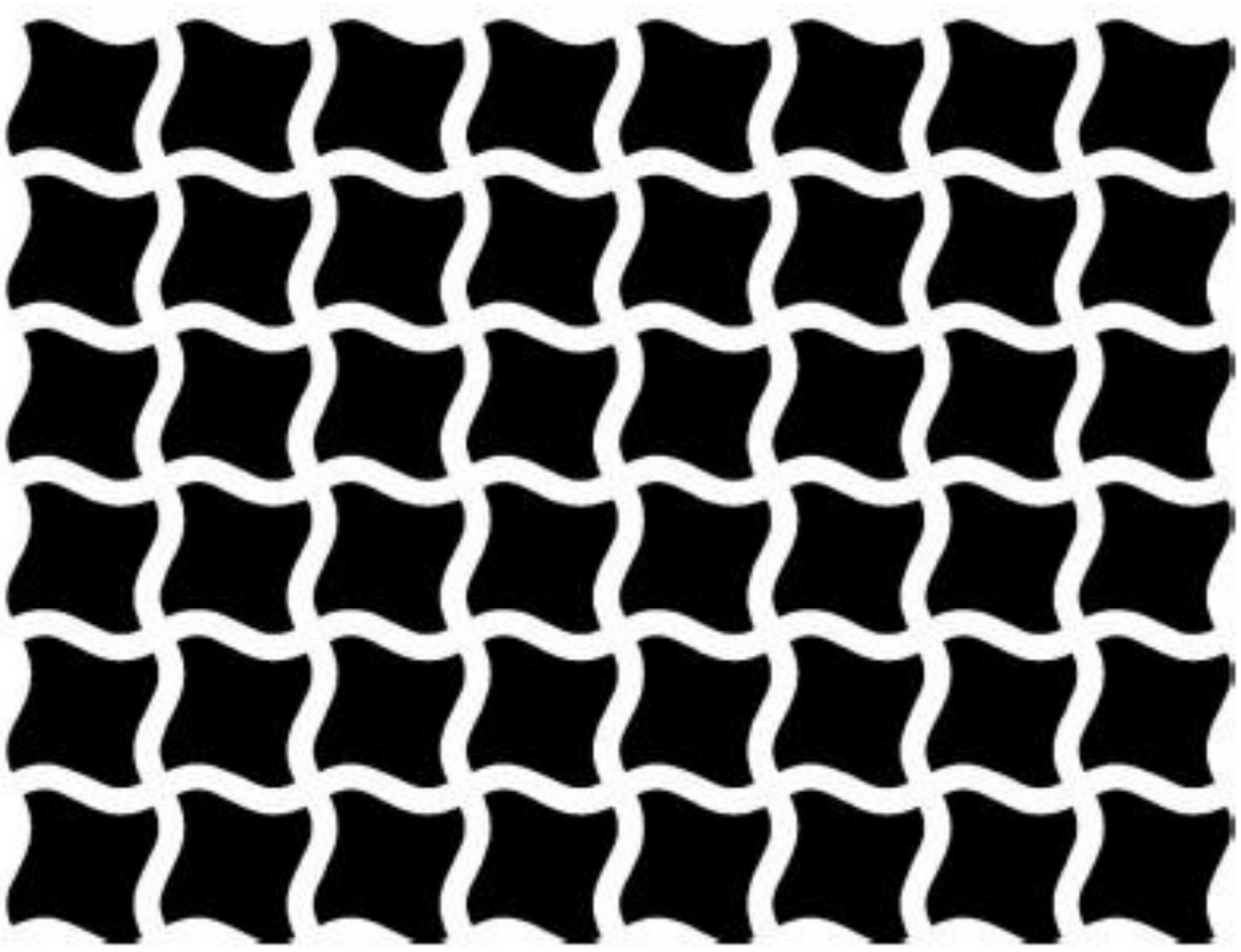

Figure 3: A counterexample by Geier (2008)[4].

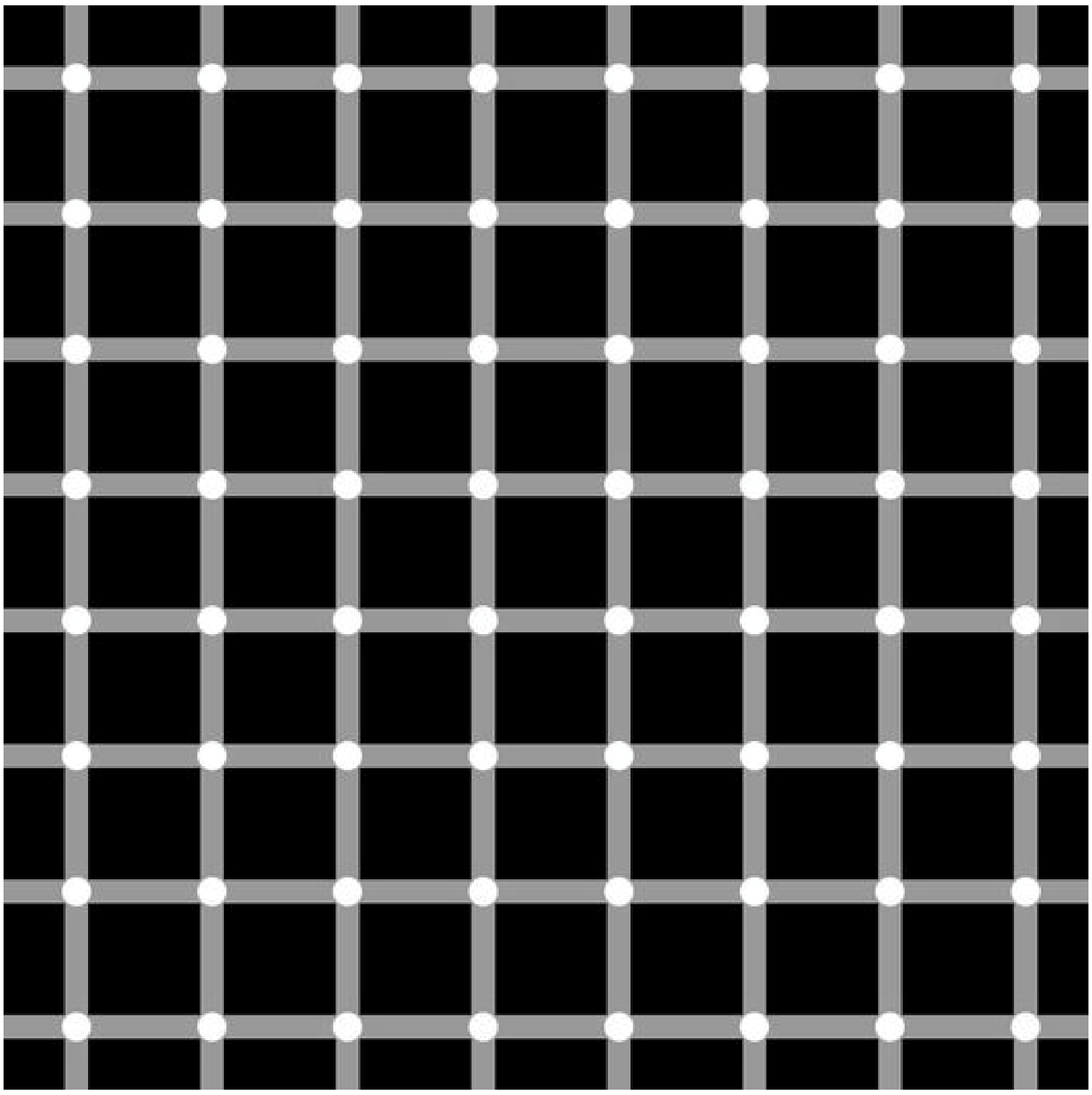

Figure 4: Schrauf's scintillating grid illusion (1995)[6].

### 3. Schrauf's scintillating grid illusion

Schrauf (1995) improved the Hermann grid illusion to create a more vivid illusion[6]. As Figure 4 shows, by making the paths gray and the intersections white circles, the illusion becomes more pronounced—this is called the scintillating grid illusion. Let's use this figure to consider what produces this illusion.

The white circles at the intersections are arranged in 8 rows and 8 columns, totaling 64 circles at equal intervals. When focusing one's gaze on the center of Figure 4 and peripherally observing the surrounding white circles, they appear black. By then moving the gaze to a black circle, it changes back to white, creating an effect as if the black circles are flashing. It can also seem as if small black dots are appearing within the white circles. These flashing black dots seem different from the black of the square background, appearing almost like black holes. I can see the illusion when using one eye, but it is more vivid when viewed with both eyes. After creating several patterns and experimenting, I developed a hypothesis, which is the main topic of this article.

## 4. Minimal example of the grid illusion

When I look at all 64 white circles in Figure 4, I simply feel puzzled and inexplicably become tired. I thus decided to extract just one row or one column of 8 circles from Figure 4 and observe that (Figures 5 and 6).

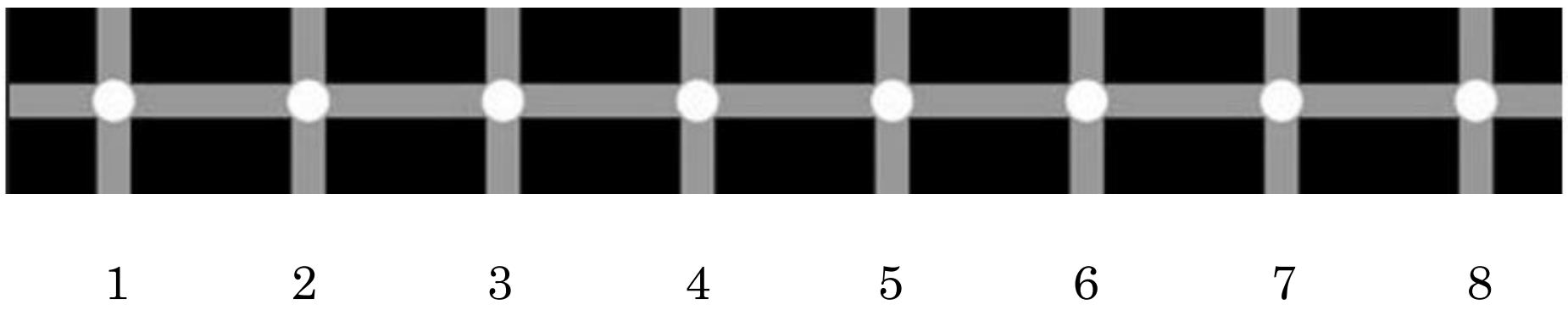

Figure 5: A single row (eight dots) from the grid illusion.

In Figure 5, shifting one’s visual focus to the left or right makes the white circles change to black dots, resulting in a flashing effect. The eight white circles are arranged horizontally, and I’ve numbered them 1 to 8 from left to right for explanatory purposes. Focusing on the central white circles 4 or 5 with both eyes makes 3 and 6 appear as white circles, but those at either end (1, 2, 7, and 8) become flashing black dots. Focusing on 1 or 2 to “chase” the black dots reverts them to white circles, but then 4 and 5 flash as black dots. Similarly, shifting focus to 7 or 8 makes the area around 4 and 5 flash.

In Figure 6, shifting visual focus up and down changes hardly any white circles into black dots. However, one must be careful to maintain a top-to-bottom focal shift, ensuring that the left and right eyes do not glance horizontally.

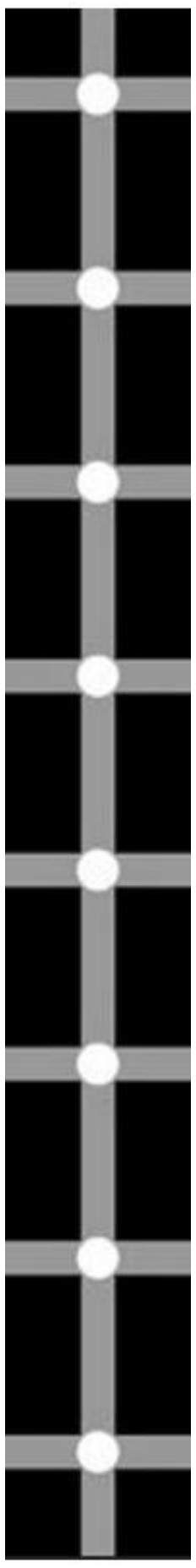

Figure 6: A single column (eight dots) from the grid illusion.

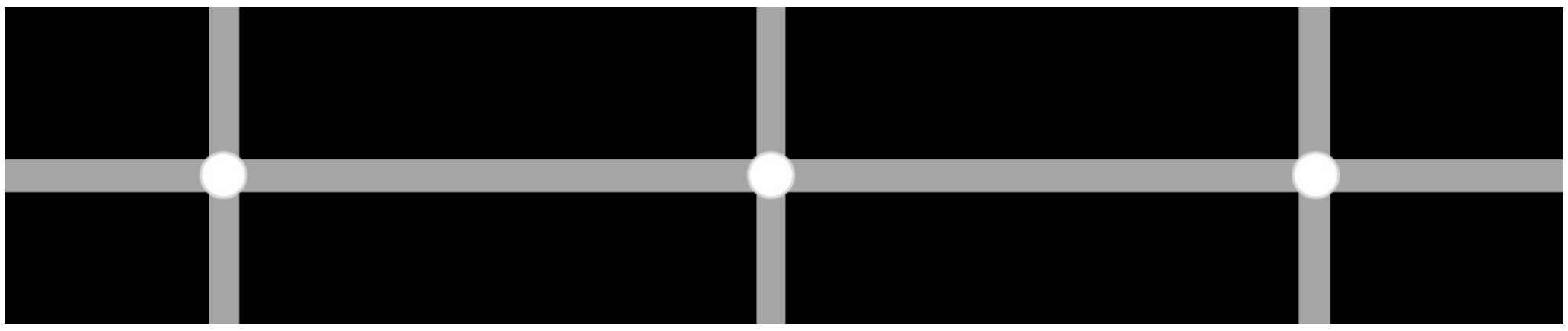

Figure 7: Minimal (three-circle) model of the grid illusion.

Considering the above, we can see that from their cause, both the Hermann grid illusion and Schrauf's scintillating grid illusion can be reproduced in one dimension rather than two. I thus developed the minimal model shown in Figure 7. This model contains only 3 white circles, further reducing the number of elements from the eight-circle horizontal row in Figure 5. Since enlarging and cropping the eight-circle model in Figure 5 would increase the width of the gray paths and the size of the white circles at intersections, I adjusted the path width and circle size in this new model.

Here, focusing on the central white circle with both eyes causes the circles on the left and right to appear as flashing black dots. This occurs due to subtle horizontal movements of the visual focus. Shifting focus to the leftmost circle, the two circles on the right flash as black dots. Conversely, focusing on the rightmost circle causes the two circles on the left to flash as black dots. After confirming this illusion several times with the minimal model in Figure 7, I began to consider that the reason for this illusion might be related to human blind spots.

## 5. Dynamic movement of blind spots

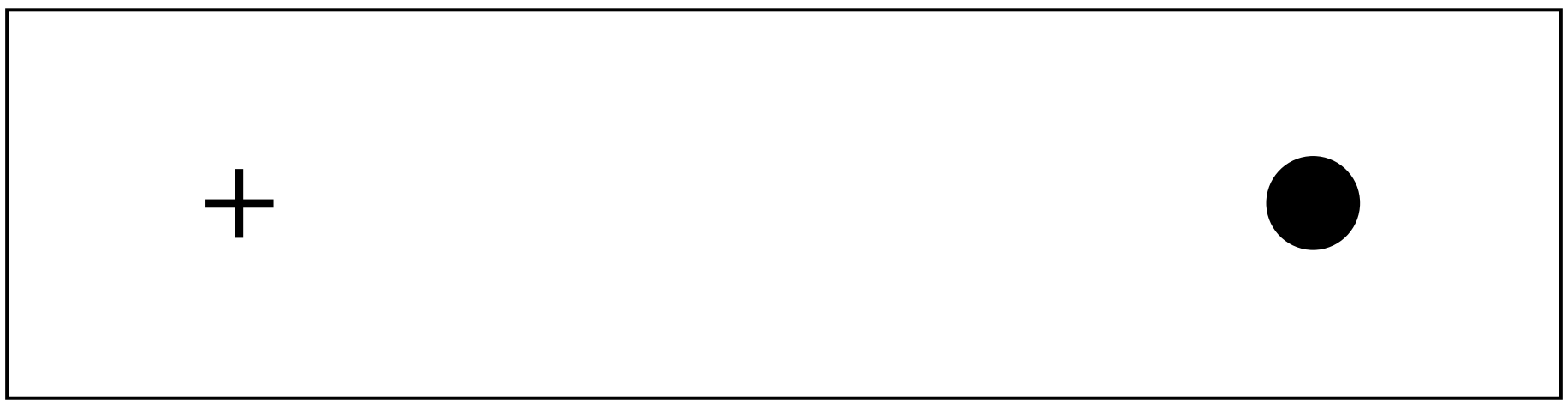

Figure 8: A blind spot experiment.

Let's try a blind spot experiment. The cross and the black dot in Figure 8 should be printed about 10 cm apart. This can also be experienced on a computer screen.

1. Hold the printed paper in your hand, bringing the center of the paper to about the level of your brow.
2. Close your left eye and look at the cross with your right eye.
3. Keeping your right eye on the cross and the paper's position fixed, move your head back and forth to find the orientation at which the black dot disappears. When the dot vanishes, it is being projected onto the blind spot of your right eye. While there are individual variations, the blind spot can typically be confirmed when the distance between the eye and the paper is about 25 to 30 cm.

Humans convert light entering the eye into electrical signals using photoreceptor cells in the retina, then process these signals in the brain to achieve final perception. This information travels to the brain via the optic nerve. At the optic nerve's exit point (i.e., the optic disc), there are no photoreceptor cells, so light cannot be sensed there. We call this area the "blind spot" or "Mariotte's spot," named after the seventeenth-century French physicist Edme Mariotte (1620–1684) who discovered it.

Figure 9 shows a schematic diagram of the human eye, as a horizontal cross-section of the right eye viewed from above. The left side is toward the nose, and the right side is toward the ear. Light passes through the lens and vitreous body to be projected onto the retina. Because the lens is convex, the image on the retina is inverted. The fovea is at the bottom. The optic disc (blind spot) is to the left of the fovea, on the nasal side.

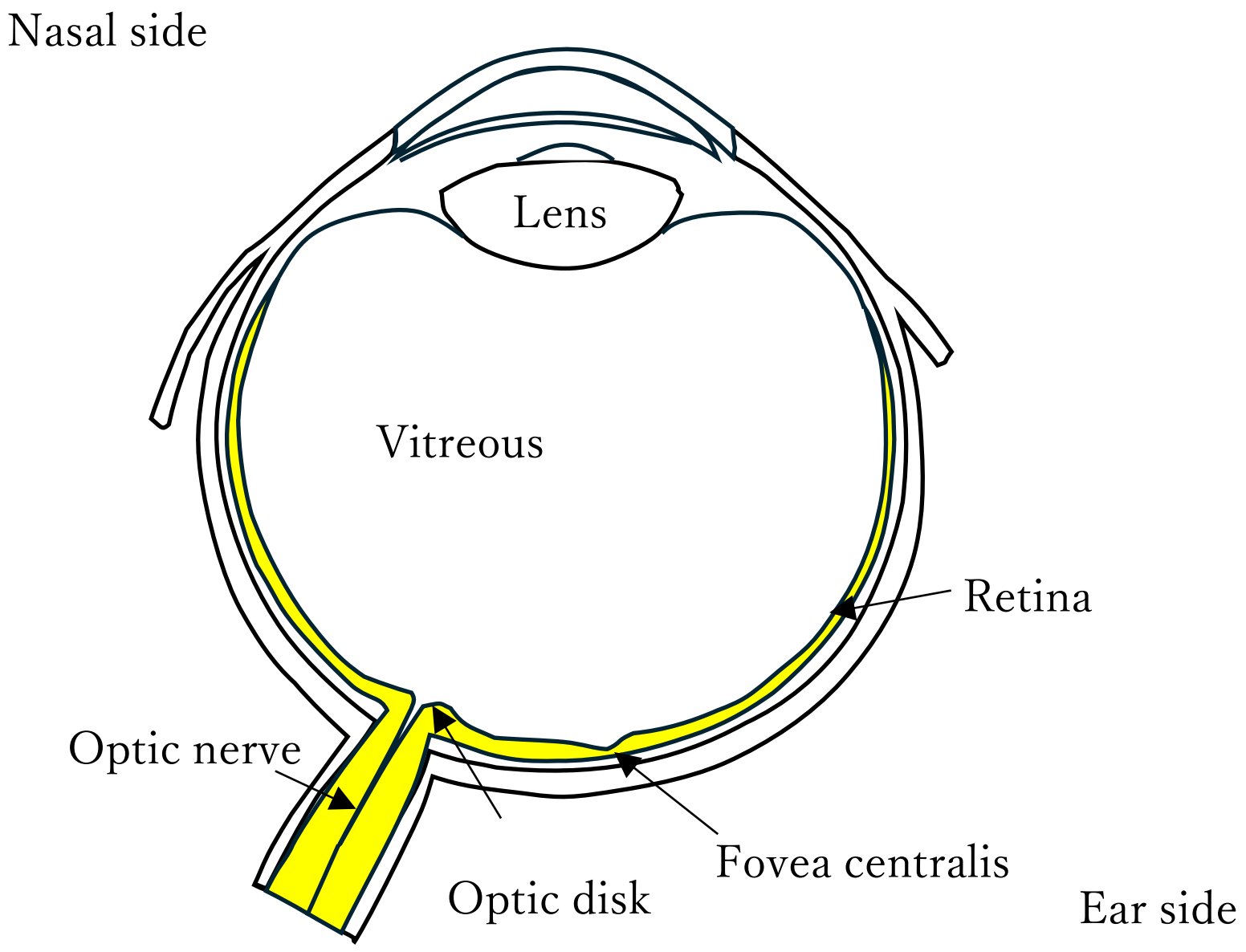

Figure 9: Cross-section of the right human eye.

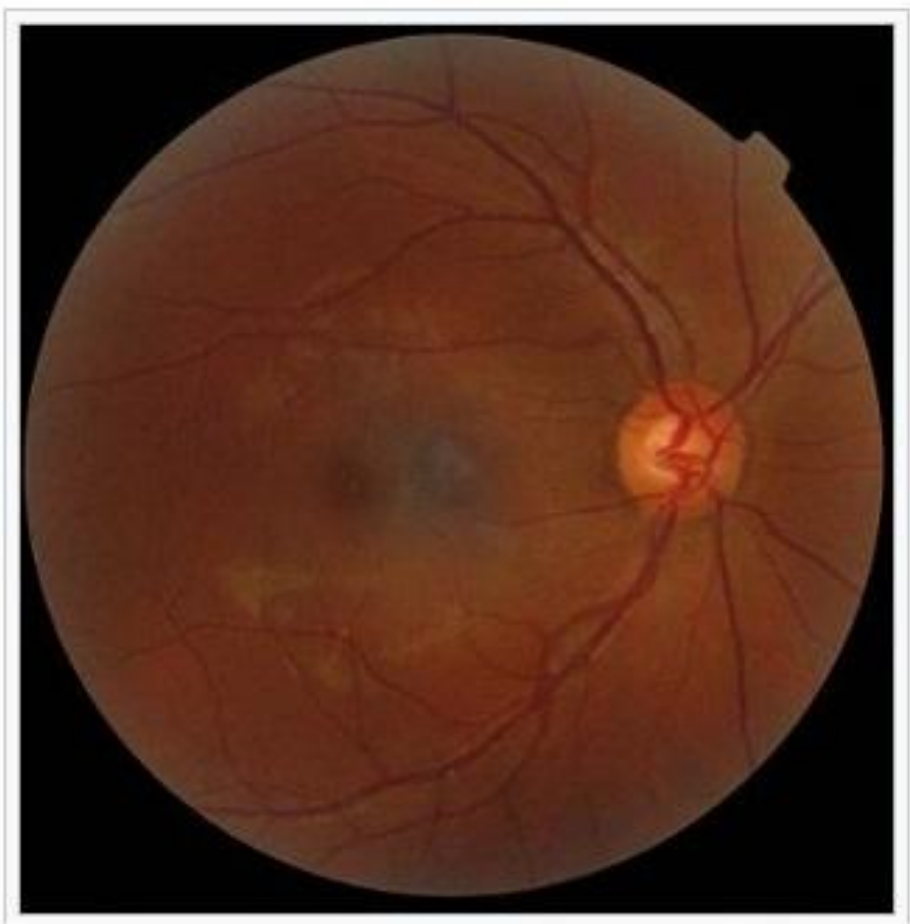

Figure 10: Optic disk (blind spot) (Source: Wikipedia[8)]).

Figure 10 is a fundus photograph of the human eye[8)]. The bright circular part is the optic disc (blind spot), which has a diameter of about 0.7 mm. In this figure, we can see the central retinal artery branching out from the optic disc to various parts of the retina. The darker area in the center is the fovea.

The blind spot is located about 15 degrees and 5 mm nasal to the fovea, which is at the center of the retina. It subtends a visual angle of about 5 degrees and has a nearly circular elliptical shape with its long axis vertical. A 5-degree visual angle is equivalent to an 8-cm-diameter circle viewed from 1 m away. Mariotte is credited as the first to measure the blind spot's size. He estimated the ratio of object size to distance from the eye as 1/9 to 1/10, meaning an object 2 cm in size would be clearly visible at a distance of 18 to 20 cm[7)].

Each eye has one blind spot, and light from a single point in the external world never enters both simultaneously. When using both eyes, objects corresponding to one eye's blind spot can be seen by the other eye, so we usually do not notice our blind spots.

Let's consider how the illusion occurs by overlaying the blind spots of both eyes onto the minimal model of the grid illusion (Figure 7), as shown in Figure 11.

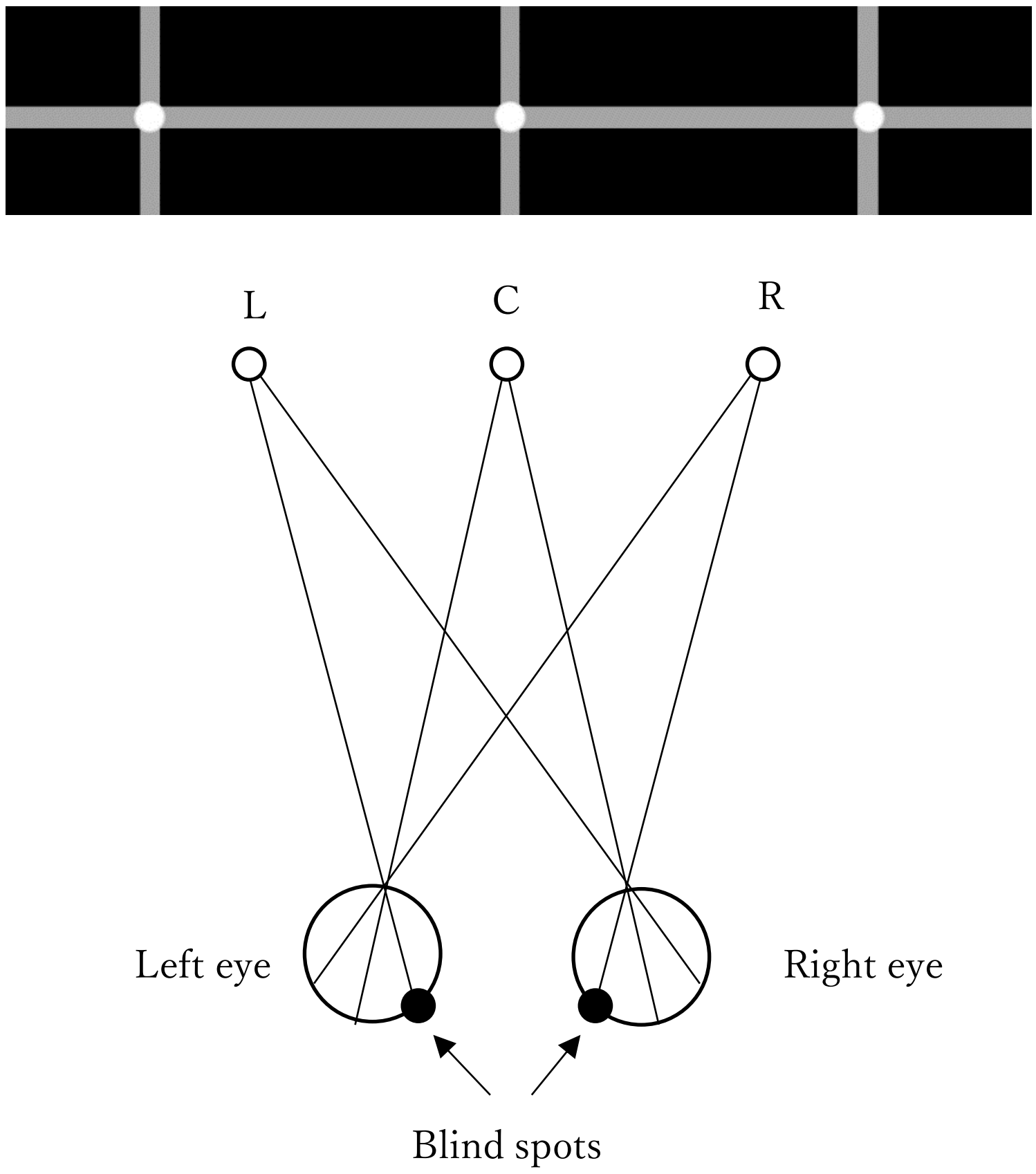

Figure 11: Blind spots in each eye.

A blind spot (optic disc) is present in both eyes. The right eye's blind spot is to the left of the fovea, while the left eye's blind spot is to the fovea's right. In other words, they are on the nasal side, not the ear side.

In the figure, three white circles are arranged in a row. Let's call them L (left), C (center), and R (right).

1. When focusing on C with both eyes, it appears as a white circle, unaffected by the blind spots.
2. When focusing on L with the right eye, it appears white, but R falls on a blind spot and thus appears as a black dot.
3. When focusing on R with the left eye, it appears white, but L falls on a blind spot and thus appears as a black dot.

Humans perform these three actions simultaneously and continuously with both eyes because our gaze is in constant motion. As our focus point moves, our blind spots move along with it, and this, I believe, results in the Hermann grid illusion.